# Cohort profile: the Northwest China Real-world and Population-based Cohort

Qi Huang [1†], Yanjun Li [1†], Bo Yin [2,3], Yaoguo Wang [4], Yujuan Yuan [5,6], Yanying Guo [7], Kuiying Gu [8], Yining Yang [5,6,9,10,*], Qian Di [1,11,*]

[1] Vanke School of Public Health, Tsinghua University, Beijing, China.

[2] School of Medicine, Tsinghua University, Beijing, China.

[3] School of Biomedical Engineering, Tsinghua University, Beijing, China.

[4] Information Center, People's Hospital of Xinjiang Uygur Autonomous Region, Urumqi, China.

[5] Department of Cardiology, People's Hospital of Xinjiang Uyghur Autonomous Region, Urumqi, China.

[6] Xinjiang Key Laboratory of Cardiovascular Homeostasis and Regeneration Research, Urumqi, China.

[7] Department of Endocrinology, People's Hospital of Xinjiang Uygur Autonomous Region, Xinjiang Clinical Research Center for Diabetes, Urumqi, China.

[8] School of Public Health, Soochow University, Suzhou, China.

[9] State Key Laboratory of Pathogenesis, Prevention and Treatment of High Incidence Diseases in Central Asia, Xinjiang Medical University, Urumqi, China.

[10] Key Laboratory of Cardiovascular Disease Research, First Affiliated Hospital of Xinjiang Medical University, Urumqi, China.

[11] Institute for Healthy China, Tsinghua University, Beijing, China.

† Joint first authors

*Corresponding author:

Yining Yang, Department of Cardiology, People's Hospital of Xinjiang Uyghur Autonomous Region, Urumqi, China. Email: yangyn5126@163.com;

Qian Di, Vanke School of Public Health, Tsinghua University, Beijing, China. Email: qiandi@tsinghua.edu.cn.

**Key Features:**

- The Northwest China Real-World and Population-based cohort is an ongoing prospective cohort with more than 25 million population, covering almost all residents across approximately 1.66 million square kilometers in northwest China;
- The cohort integrates data from various sources, including health profiles, examination records, electronic health records, mortality records, statistical yearbooks, and environmental datasets, covering comprehensive health-related factors such as demographics, lifestyle factors, family medical history, living conditions, enrollment in national public health services, physical examinations, blood assay tests, diagnostic assessments, disease outcomes, and cause-specific mortality.
- This real-world dataset can evaluate clinical treatment effectiveness and prognosis, assess impact of health policy, and investigate the health effects of multiple risk factors .
- From January 2019 to December 2023, the cohort has included 13,634,481 participants, accumulating 47,050,707 person-years of follow-up, with 13,598,407 medical diagnosis records and 881,114 recorded deaths.
- Cohort data are available upon request. De-identified and anonymized data are stored on local servers and accessed through a data-sharing platform, enabling users to utilize the data without direct access to the raw information. A description of the proposed research can be sent to Yining Yang & Qian Di.

- **Why was the cohort set up?**

**Background**

The global burden of chronic diseases is increasing due to aging populations and environmental changes, with disability-adjusted life-years (DALYs) rising from 2.63 billion in 2010 to 2.88 billion in 2021.[1] Chronic diseases impose growing health burden on society in China: in 2017, chronic conditions such as stroke and ischemic heart disease accounted for 20 of the top 25 diseases and contributed 86.39% of the total burden.[2] Building cohorts to study chronic diseases is essential for identifying risk factors and developing solutions.

While traditional cohort construction is time-consuming and expensive, using existing real-world data, such as electronic health records, provides a more efficient, cost-effective way to create large-scale cohorts and improve research outcomes.[3-5] Real-world data (RWD) refers to health-related data collected from routine clinical practice, such as electronic health records, insurance claims, or patient registries.[6] To construct a cohort using real-world data (RWD), researchers should carefully define baseline characteristics, specify the exposure, determine outcome measures, and establish appropriate follow-up periods, thereby minimizing biases and enhancing causal inference[7] This method reduces costs and expands population coverage, improving research validity. For example, in Denmark, national databases like the Danish National Patient Registry and the Education Registry have been used to build real-word and population based cohort, for study topics such as health inequalities[3], air pollution[8], adverse health outcomes[4], and behaviors[9]. Additionally, a similar approach is taken in Europe[10] and the U.S.[5,11], but China lacks large-scale cohort studies using real-world data. This study aims to utilize existing digital infrastructure in a northwestern region of China to build a cohort more efficiently, avoiding the high costs of traditional methods.

Second, cohort studies should incorporate a comprehensive range of disease risk factors, including environmental, lifestyle, behavioral, and metabolic factors, and include multiple health endpoints, especially given the increasing prevalence of comorbidity and multimorbidity, with over 50% of adults above 60 globally having multiple chronic conditions, a rate that has risen over the past two decades.[12,13] Comorbidities are common in diseases such as cardiovascular disease[14], diabetes[15], chronic obstructive pulmonary disease[16], asthma[17], mental disorder[18], especially among some sub-populations, such as among those previously infected with COVID-19[19]. Understanding the factors associated with comorbidity is essential, as it is a complex issue that intensifies with age and varies across different morbidity profiles.[12,15] Take metabolic risk factors as an example, from 1990 to 2021, metabolic risks have increased by 49.4% (95% CI: 42.3–56.9%), surpassing environmental and occupational risks to become the second most substantial factor in global disease burden.[20] Metabolic risks like high systolic blood pressure, high body mass index (BMI), high fasting plasma glucose (FPG), and high LDL cholesterol significantly impact older adults' health. Specifically, global age-standardized

DALY rates attributable to high BMI and high FPG have risen by 15.7% (9.9–21.7) and 7.9% (3.3–12.9%), respectively, over the past two decades, with exposures increasing annually by 1.8% (1.6–1.9) for high BMI and 1.3% (1.1–1.5) for high FPG. [20] These trends highlight the critical need to focus on these risk factors in research, alongside other influential factors such as lifestyle and environmental elements.

Third, understanding the real-world impacts of policy exposure is also critical, especially in the realm of chronic disease control. Over the years, a multitude of interventions have been launched to control chronic diseases. Research has repeatedly underscored the efficacy of lifestyle interventions: diet and exercise interventions can reduce diabetes incidence among people with impaired glucose tolerance[21], lower cardiovascular disease incidence and mortality[22]. Lifestyle intervention is more effective than drug use to prevent diabetes[23]. Beyond individual-level interventions, city-wide strategies, encompassing community-wide health education initiatives, have achieved improvements in cholesterol levels, blood pressure, and resting pulse rates, ultimately curtailing overall mortality[24]. Consequently, national-level intervention policies have been implemented to control chronic diseases. In China, the government launched the National Basic Public Health Services Programme (NBPHSP) in 2009, which is fully funded by the government through performance-based capitation payments made in two installments each fiscal year.[25] This program includes creating health profiles for each resident and enrolling individuals with hypertension and diabetes in chronic disease management programs, among other services. In the United States, the Center for Disease Control and Prevention also offers the 10 Essential Public Health Services, with special emphasizes on chronic disease prevention.[26] However, despite of wide application of such national-scale chronic disease prevention programs, few empirical studies have carried out to evaluate the effectiveness of such prevention programs.

To address above needs, this study constructed a large-scale and ongoing prospective cohort based on the real-world data in northwestern China. The cohort integrates data from various sources, including health profiles, examination records, electronic health records, mortality records, statistical yearbooks, and environmental datasets. The cohort included comprehensive health-related factors such as demographics, lifestyle factors, family medical history, living conditions, enrollment in national public health services, physical examinations, blood assay tests, diagnostic assessments, disease outcomes, and cause-specific mortality. Thus, we can build a cohort that cover the entire regional population that include multiple health endpoints, comprehensive disease risk factors and real-world information on policy interventions.

- **Who is in the cohort?**
  - **Cohort overview**

Our population-based cohorts included almost all residents of northwestern China with community healthcare center registers files, approximately 18 million individuals. This cohort was established using data from 12,520 primary healthcare institutions, including township health centers, village clinics, and community health service centers. The baseline was formed from resident health profiles, which included demographic information and tracked participants' status as active, migrated, or deceased, similar to Denmark's register-based cohort study.[3] Follow-up information is primarily obtained from chronic disease management programs, annual health examinations, and supplemented by hospitalization records. Cause-specific mortality records from local death registrations. All above information from multiple sources is linked together using a person's encrypted national ID number to construct our cohort. Finally, the cohort has included 13,634,481 participants, accumulating 47,050,707 person-years of follow-up, with 13,598,407 medical diagnosis records and 881,114 recorded deaths (Figure 1).

- **Enrollment**

Our enrollment phase spanned from Jan 2019 to Dec 2023. Local residents were selected if they met the following criteria: (1) with available health profiles at local community healthcare centers after Jan 2019; (2) with ages ≥ 18 years. We excluded individuals (1) with duplicates (2) with missing or wrong ID numbers; (2) with missing or wrong key information, such as age and sex. By the end of our enrollment process, 17,238,412 individuals were included at baseline. The average age of the included population is 42 years, with males comprising 49.54% and females 50.46%. Descriptive Information on educational level, occupation, marital status, family history of diseases, medical insurance, blood type, and household living conditions (including type of fuel used, drinking water source, kitchen ventilation, and type of toilet) is presented in Table 1.

Based on this population cohort, we can construct several disease-specific cohorts. Constructing a disease-specific cohort requires selecting a population that is disease-free at baseline. To achieve this, individuals with pre-existing conditions, as documented in their health profiles and electronic health records, can be excluded. For instance, when constructing a cardiovascular disease cohort to investigate the impact of heatwaves on cardiovascular health, we would first exclude individuals who had the disease at baseline, based on their initial registration records and medical history prior to cohort entry. Once the baseline population is defined, environmental factors (i.e. heat exposure) is matched geographically, followed by tracking health outcomes through annual health exam and medical records. Thus, this 17-million-cohort based on real-world data enables the construction of multiple disease-specific cohorts, empowering researchers to conduct in-depth analyses of disease patterns, identify risk factors, and evaluate the effectiveness of interventions.

- **How often have they been followed up?**

After enrollment, follow-up information was collected by scheduled and event-driven follow-up. Annual health exam project provided by 7,523 primary centers across the region provides scheduled follow-up of the study population. This project's follow-up collects information in five categories: general health status, lifestyle, physical examinations, blood biochemical tests, and instrumental examinations (Table 2). Each year, over 9 million individuals are followed, amounting to a total of 47.05 million follow-up visits over five years. The cohort comprises 14.00 million participants who had at least one health examination, 11.00 million with at least two, 9.13 million with at least three, 7.32 million with at least four, and 5.05 million with at least five health exams.

Secondly, event-driven follow-up information extracted from 163 hospitals and 63 Traditional Chinese Medicine(TCM) hospitals, covering all tertiary hospitals in the region. More than 17 million inpatients records were collected from Jan 2016 to Dec 2023, including 13.60 million records from the cohort population. Inpatients records collected participants' basic information, disease diagnostic information (Table 2). All disease diagnoses were based on the International Classification of Diseases, 10th Revision (ICD-10) codes. In total, there are 1.28, 1.52, 1.81, 2.04, 3.22 inpatients records in the cohort participants, from 2019 to 2023, respectively.

Moreover, data on cohort population migration and death are collected by primary healthcare personnel, including the time of migration, time of death, and specific causes of death. In this study, we consider both migration and death as right-censored events. Migration to areas outside the study region significantly impacts our ability to collect outcome information, representing a crucial reason for loss to follow-up. Thus, in this cohort, we consider both migration and death as right-censored events. In total, 2.72 million participants (14.47%) were lost to follow-up over the five-year period, and 0.88 million deaths were recorded.

- **What has been measured?**

  **Baseline measurements**

The cohort's baseline information is derived from residents' health profiles, which includes demographic information such as sex, age, marital status, education, and blood type; medical history, family history, genetic history, and disability status; as well as living environment details like type of fuel used, drinking water source, and type of toilet (Table 2).

Residents' health profiles were established mainly through two methods. First, when they sought medical services at primary healthcare centers, such as township health centers, village clinics, or community health service centers, healthcare professionals created their health profiles. Second, their profiles were also compiled through home visits, disease screenings, and health examinations conducted by primary healthcare teams. These profiles were digitized and stored in local electronic health record data centers. Residents' health records are subsequently retrieved and updated each time they actively seek medical services or participate in home visits, disease screenings, and health examinations. [27]

**Annual Health Exam Project**

Individuals with established health profiles can receive annual health exam at local healthcare institutions free of charge. The health exam includes general health examinations, lifestyle assessments, health status evaluations, biochemical tests, instrumental examinations, and health evaluation and guidance (Table 2). All items from the health exam can be utilized as covariates in the cohort.

General health examinations include measurements of temperature, pulse rate, blood pressure, height, weight, waist circumference, and assessments of self-care ability, cognitive function, and emotional state in the elderly. Self-care ability in the elderly is evaluated in five areas: eating, grooming, dressing, toileting, and mobility. Scores are summed to classify individuals as follows: 0-3 points indicate self-sufficiency, 4-8 points indicate mild dependency, 9-18 points indicate moderate dependency, and ≥19 points indicate inability to self-care. Cognitive function and emotional state in the elderly are initially screened through questions, with those screening positive requiring further evaluation using the Mini-Mental State Examination (MMSE) and the Geriatric Depression Scale (GDS). Lifestyle assessments cover physical exercise, dietary habits, smoking and alcohol consumption, and occupational exposure to harmful factors. Health status evaluations include organ examinations and specific physical examination items (Table 2). Auxiliary examinations encompass biochemical tests of blood and urine, as well as instrumental examinations such as electrocardiograms (ECG), X-rays, and ultrasounds. Blood tests include complete blood counts (hemoglobin concentration, white blood cell count, platelet count) using automated blood cell analyzers. Liver and kidney function tests measure serum alanine aminotransferase (ALT), aspartate aminotransferase (AST), albumin, serum creatinine, blood urea, blood potassium concentration, and blood sodium concentration using automated biochemical analyzers. Urinalysis includes tests for protein, glucose, ketones, and occult blood in the urine. Fasting blood glucose is measured using the glucose oxidase method (GOD-POD method) with an automated biochemical analyzer. The health assessment section involves physicians providing health evaluations and relevant health guidance based on the specific results of the participant's health examination. Details were showed in supplementary 1-4.

**Inpatients and mortality records**

All inpatient records for individuals with established health records were collected, primarily including basic information, disease information, and diagnostic information. Basic information includes age, sex, blood type, home address, and drug allergy history. Disease information includes admission time, discharge time, admitting hospital, and mode of discharge (transfer, discharge, death). Diagnostic information includes outpatient diagnosis, inpatient diagnosis, discharge diagnosis, pathological diagnosis, and codes for external causes of injury and poisoning. All disease diagnoses were coded by the ICD-10. Inpatient information was documented by qualified medical personnel and entered into an electronic health information system. We aggregated the hospital information and matched it to the health record data using each individual's national unique identification ID. Death information, including death time and specific cause, is mainly obtained through hospital and primary centers.

**Environmental factor measurements**

We have collected gridded environmental data, including monthly average temperature[28], monthly maximum temperature[29], monthly minimum temperature[29], $PM_{2.5}$ levels[30], $PM_{2.5}$ components[31] ($SO_4^{2-}$, $NO_3^-$, $NH_4^+$, BC, OM), $O_3$ [32], and nighttime light data[33]. We matched the latitude and longitude coordinates to the cohort based on the residential addresses of the cohort population, and then matched the environmental data to the cohort based on these coordinates.

- **What has it found?**

This cohort offers a valuable dataset for investigating comorbidities, disease risk factors, the effectiveness of policy interventions, etc. As an example, we examined the annual health program, an integral component of China's National Basic Public Health Services, and its association with key health outcomes, including mortality rates and hospitalization duration. We established a cohort utilizing residents' health records as the baseline, with participation in the previous year's annual health examination as the exposure variable, and both mortality and length of hospital stay as outcome measures. Cox proportional hazards models were employed to assess the effect of health exam on mortality, and linear mixed-effects models were used to evaluate their impact on hospitalization duration.

In total, 13,634,481 individuals with 47,050,707 followed person-years were included in the study. The average age of participants was 41±16.12 years, with 49.54% being female. The annual mortality rate was 4.66‰, and the median hospital stay was 7 days. Individuals who participated in the annual health examination in the previous year had a lower risk of mortality in the current year compared to those who did not (HR: 0.713, 95% CI: 0.707-0.719). Additionally, the cumulative number of health exam up to the previous year was negatively associated with current-year mortality risk (HR: 0.895, 95% CI: 0.892-0.898), indicating that each additional examination was associated with a 10.5% reduction in mortality risk. Moreover, after adjusting for other factors and individual differences, each additional examination was associated with a reduction of approximately 1.127 (95% CI: 1.124-1.129) hospital days on average. According to reports from National Health Commission of the Peoples Republic of China, the average daily hospitalization cost per person is 1,551 CNY [34,35], implying that each additional health examination reduces subsequent hospital stays by 1.127 days, resulting in a cost saving of 1,748 CNY.

In summary, this study provides evidence of the health and economic benefits of China's health examination program, including possible reductions in hospitalization duration and costs, lower mortality risk, and the prevention of premature deaths. These findings suggest the need for continued refinement of public health policies and improvements in clinical practice.

- **What are the main strengths and weaknesses?**

The first limitation of this study is the lack of comprehensiveness in the baseline survey data, which does not include omics data such as genomic and proteomic information, blood biomarkers, and results from instrumental examinations. However, relevant biochemical blood test data will be available from the subsequent five years of health examination information, allowing for the adjustment of baseline data in accordance with the research objectives. Second, the scope of the current dataset is relatively limited, incorporating basic public health records, health examination data, and hospital admission information, supplemented by publicly available environmental data. Ongoing efforts are focused on broadening the dataset by integrating electronic health records, disease surveillance, mortality surveillance, and ecological data from multiple centers and time points.

Nevertheless, the study has significant strengths. First, this cohort covers a geographical area of 1.66 million square kilometers, includes over 99% of the regional population, and has a sample size exceeding 17 million, with nearly 10 million participants in routine health examinations annually. Second, the cohort integrates data from multiple sources and covers a wide range of health-related factors such as demographics, lifestyle, family medical history, living conditions, public health service enrollment, physical examinations, blood tests, diagnostic assessments, disease outcomes, and cause-specific mortality. Finally, the study achieves an optimal balance between data sharing and privacy protection, and data security. De-identified and anonymized data are stored on local servers and accessed through a data-sharing platform, enabling users to utilize the data without direct access to the raw information.

- **Can I get hold of the data? Where can I find out more?**

We welcome collaboration from over the world to maximize the usage of this population-based multi-disease cohorts. We have uploaded part of the data to the server, which users can access and analyze via the internet. The user interface, as shown in Fig. 2, allows users to customize the selection of population covariates and statistical models for analysis. For permission requests, please contact Yining Yang & Qian Di

- **Ethics approval**

The Tsinghua University Science and Technology Ethics Committee (Medicine) has approved this study (Project No: THU01-20240123).

- **Author contributions**

Qi Huang: data cleaning, coding, cohort design, analysis, and drafting. Yanjun Li: data cleaning, coding, and analysis. Bo Yin: coding and environmental variable matching. Yaoguo Wang and Yujuan Yuan: data management. Ningyi Yang: data provision, funding, and final review. Qian Di: conceptualization, manuscript revision, funding, and final approval.

- **Supplementary data**

Supplementary data are available at IJE online.

- **Funding**

- **Conflict of interest**

The authors declare that they have no known competing financial interests or personal relationships that could have influenced the work reported in this paper.

- **References**

## Table 1. Characters of baseline population.

| Characters | Male (N= 8697874) | Female (N= 8540538) | Overall (N= 17238412) |
|---|---|---|---|
| **Age, (years)** [a] | | | |
| Mean (Standard Deviation) | 42.30 (16.13) | 42.32 (16.11) | 42.31 (16.12) |
| Median [25%, 75%] | 41.0 [29.0, 52.0] | 41.0[29.0, 52.0] | 41.0 [29.0, 52.0] |
| **Occupation, n (%)** [a] | | | |
| Civil Servant | 487556 (5.6) | 418835 (4.9) | 906391 (5.3) |
| Professional and technical personnel | 213618 (2.5) | 208434 (2.4) | 422052 (2.4) |
| Clerical and related staff | 4261695 (49.0) | 4058522 (47.5) | 8320217 (48.3) |
| Service and sales worker | 502179 (5.8) | 499124 (5.8) | 1001303 (5.8) |
| Farmer | 331347 (3.8) | 196886 (2.3) | 528233 (3.1) |
| Production and transportation worker | 385393 (4.4) | 393012 (4.6) | 778405 (4.5) |
| Other unclassified workers | 1635115 (18.8) | 1739151 (20.4) | 3374266 (19.6) |
| Unemployed and retired | 490498 (5.6) | 676265 (7.9) | 1166763 (6.8) |
| Unknown | 390473 (4.5) | 350309 (4.1) | 740782 (4.3) |
| **Education, n (%)** [b] | | | |
| Postgraduate | 20058 (0.2) | 25745 (0.3) | 45803 (0.3) |
| Bachelor's degree | 278306 (3.2) | 319583 (3.7) | 597889 (3.5) |
| Junior college education | 973186 (11.2) | 1050417 (12.3) | 2023603 (11.7) |
| Technical school | 60110 (0.7) | 42908 (0.5) | 103018 (0.6) |
| Secondary vocational education | 891662 (10.3) | 822349 (9.6) | 1714011 (9.9) |
| General higher education | 490395 (5.6) | 455005 (5.3) | 945400 (5.5) |
| General primary education | 3226489 (37.1) | 2885166 (33.8) | 6111655 (35.5) |
| Primary school education | 2061147 (23.7) | 2189789 (25.6) | 4250936 (24.7) |
| Illiterate or semi-literate | 238973 (2.7) | 362496 (4.2) | 601469 (3.5) |
| Unknown | 457548 (5.3) | 387080 (4.5) | 844628 (4.9) |
| **Marital status, n (%)** [a] | | | |
| Unmarried | 1443224 (16.6) | 890244 (10.4) | 2333468 (13.5) |
| Married | 6838797 (78.6) | 6919693 (81.0) | 13758490 (79.8) |
| Divorce | 153191 (1.8) | 165577 (1.9) | 318768 (1.8) |
| Widow | 135707 (1.6) | 460136 (5.4) | 595843 (3.5) |
| Unknown | 126955 (1.5) | 104888 (1.2) | 231843 (1.3) |
| **Family history of diseases, n (%)** | | | |
| Yes | 65365 (0.8) | 72038 (0.8) | 137403 (0.8) |
| No | 8296586 (95.4) | 8179336 (95.8) | 16475922 (95.6) |
| Unknown | 335923 (3.9) | 289164 (3.4) | 625087 (3.6) |
| **Family history of diseases, n (%)** | | | |
| Yes | 109130 (1.3) | 203791 (2.4) | 312921 (1.8) |
| No | 8588744 (98.7) | 8336747 (97.6) | 16925491 (98.2) |
| **Residence type, n (%)** | | | |
| Registered | 7629020 (87.7) | 7556494 (88.5) | 15185514 (88.1) |
| Non-registered | 1054541 (12.1) | 971428 (11.4) | 2025969 (11.8) |
| Unknown | 14313 (0.2) | 12616 (0.1) | 26929 (0.2) |
| **Health insurance, n (%)** | | | |
| Rural resident medical insurance | 1553416 (17.9) | 1564003 (18.3) | 3117419 (18.1) |
| Urban resident medical insurance | 4292068 (49.3) | 4299683 (50.3) | 8591751 (49.8) |
| Urban employee medical insurance | 1553238 (17.9) | 1472136 (17.2) | 3025374 (17.6) |
| Commercial insurance | 41558 (0.5) | 41438 (0.5) | 82996 (0.5) |
| Medical assistance | 48735 (0.6) | 46823 (0.5) | 95558 (0.6) |
| Self-pay | 172524 (2.0) | 165243 (1.9) | 337767 (2.0) |
| Public insurance | 14760 (0.2) | 9944 (0.1) | 24704 (0.1) |
| Other social insurance | 1553416 (17.9) | 1564003 (18.3) | 3117419 (18.1) |
| **Blood type, n (%)** | | | |
| A | 2141280 (24.6) | 2140293 (25.1) | 4281573 (24.8) |
| AB | 846489 (9.7) | 808929 (9.5) | 1655418 (9.6) |
| B | 2625254 (30.2) | 2650355 (31.0) | 5275609 (30.6) |
| O | 2303676 (26.5) | 2293449 (26.9) | 4597125 (26.7) |
| unknown | 781175 (9.0) | 647512 (7.6) | 1428687 (8.3) |
| **Cooking fuel type, n (%)** | | | |
| Liquefied gas | 839841 (9.7) | 802230 (9.4) | 1642071 (9.5) |
| Coal | 3185066 (36.6) | 3006156 (35.2) | 6191222 (35.9) |
| Biogas | 16656 (0.2) | 15208 (0.2) | 31864 (0.2) |
| Firewood | 2268141 (26.1) | 2170377 (25.4) | 4438518 (25.7) |
| Other | 839841 (9.7) | 802230 (9.4) | 1642071 (9.5) |
| **Drinking water, n (%)** | | | |
| Tap water | 6728181 (77.4) | 6589741 (77.2) | 13317922 (77.3) |
| Purified water | 196387 (2.3) | 207060 (2.4) | 403447 (2.3) |
| Well water | 327920 (3.8) | 300923 (3.5) | 628843 (3.6) |
| River or lake water | 57588 (0.7) | 52605 (0.6) | 110193 (0.6) |
| Pond water | 13141 (0.2) | 12167 (0.1) | 25308 (0.1) |
| Other | 42335 (0.5) | 36602 (0.4) | 78937 (0.5) |
| **Kitchen ventilation facilities, n (%)** | | | |
| Ventilation fan | 316778 (3.6) | 313733 (3.7) | 630511 (3.7) |
| Chimney | 2582048 (29.7) | 2422273 (28.4) | 5004321 (29.0) |
| Range hood | 2998409 (34.5) | 3092431 (36.2) | 6090840 (35.3) |
| None | 1233623 (14.2) | 1150671 (13.5) | 2384294 (13.8) |
| Unknown | 1567016 (18.0) | 1561430 (18.3) | 3128446 (18.1) |
| **Toilet type, n (%)** | | | |
| Simple shelter toilet | 1841038 (21.2) | 1692299 (19.8) | 3533337 (20.5) |
| Open pit latrine | 266494 (3.1) | 244095 (2.9) | 510589 (3.0) |
| Flush toilet | 2496314 (28.7) | 2632107 (30.8) | 5128421 (29.8) |
| Sanitary toilet | 2195977 (25.2) | 2100982 (24.6) | 4296959 (24.9) |
| Single or double-pit latrine | 319216 (3.7) | 298426 (3.5) | 617642 (3.6) |
| Unknown | 1578835 (18.2) | 1572629 (18.4) | 3151464 (18.3) |
| **Follow-up years (years)** | | | |
| Mean (Standard Deviation) | 3.81 (1.55) | 3.95 (1.48) | 3.88 (1.52) |
| Median [25%, 75%] | 4.58 [2.93, 5.00] | 4.69 [3.46, 5.00] | 4.64 [3,22, 5.00] |

[a] Sex, occupational classification, and marital status was based on the National Standard of People's Republic of China "Classification and Codes of Basic Personal Information (GB/T 6565-2009)."

[b] Education refers to an individual's highest level of academic achievement, based on the National Standard of People's Republic of China "Codes for record of formal schooling (GB/T4658-2006)"

Table 2. Summary of collected information fields for cohort population

| |
|---|
| **Part 1. Resident health record registration** |
| 1. Personal basic information: <br> Home address; Sex; Date of birth; Ethnicity; Educational level (Postgraduate, Bachelor's degree, Associate degree or specialized college, Secondary specialized school, Technical school, High school, Junior high school, Primary school, Illiterate or semi-literate, Unknown);Occupation (Officials of state organs, party and mass organizations, enterprises, and institutions; Professional technicians; Clerks and related personnel; Commercial and service personnel; Agricultural, forestry, animal husbandry, fishery, and water conservancy production personnel; Production and transportation equipment operators and related personnel; Other occupations difficult to classify; Unemployed);Marital status (Single, Married, Widowed, Divorced, Marital status not stated);Medical expense payment method (Urban employee basic medical insurance, Urban resident basic medical insurance, New rural cooperative medical system, Poverty relief, Commercial medical insurance, Full public funding, Full self-pay, Other) |
| 2. Medical and Family history of diseases: <br> Past medical history of diseases; History of drug allergies (None, Penicillin, Sulfonamides, Other); History of exposure to risk factors (None, Chemicals, Toxins, Other); Family history of diseases (Mother, Father, Siblings, Children: None, Hypertension, Diabetes, Coronary heart disease, Chronic obstructive pulmonary disease, Malignant tumor, Stroke, Severe mental disorders, Tuberculosis, Hepatitis, Congenital malformations, Other) <br> Genetic history <br> Disability status (No disability, Visual impairment, Hearing impairment, Speech impairment, Physical disability, Intellectual disability, Mental disability, Other disabilities) |
| 3.Living environment: <br> Kitchen ventilation facilities (None, Range hood, Ventilation fan, Chimney) <br> Fuel type (Liquefied gas, Coal, Natural gas, Biogas, Firewood, Other) <br> Drinking water (Tap water, Purified/filtered water, Well water, River/lake water, Pond water, Other) <br> Toilet (Sanitary toilet, Single or double-pit latrine, Flush toilet, Open pit latrine, Simple shelter toilet) <br> Poultry/livestock pens (None, Separate, Indoor, Outdoor) |
| **Part 2. Health examination information** |
| 1.General health status: <br> Current symptoms (None, Headache, Dizziness, Palpitations, Chest tightness, Chest pain, Chronic cough, Sputum production, Shortness of breath, Excessive thirst, Frequent urination, Weight loss, Fatigue, Joint swelling and pain, Blurred vision, Numbness in hands or feet, Urgency of urination, Painful urination, Constipation, Diarrhea, Nausea and vomiting, Dizziness, Tinnitus, Breast tenderness, Other); Body temperature, Pulse rate, Respiratory rate, Left systolic blood pressure, Left diastolic blood pressure, Right systolic blood pressure, Right diastolic blood pressure, Height, Weight, Waist circumference, Body mass index |
| 2.Elderly health self-assessment: <br> Satisfaction with health (Satisfied, Basically satisfied, Unclear, Not very satisfied, |

Dissatisfied), Self-assessment of ability to perform daily activities (Independent [0–3 points], Mild dependence [4–8 points], Moderate dependence [9–18 points], Unable to perform daily activities [≥19 points]), Cognitive function, Cognitive function score, Emotional state screening result, Emotional state score

3.Lifestyle:

Exercise frequency (Daily, More than once a week, Occasionally, No exercise), Duration of each exercise session, Type of exercise, Years of consistent exercise;

Dietary habits (Balanced diet, Mainly meat, Mainly vegetarian, Salt preference, Oil preference, Sugar preference);

Smoking status (Never smoked, Quit smoking, Currently smoking), Daily smoking quantity, Age started smoking, Age quit smoking;

Alcohol consumption frequency (Never, Occasionally, Regularly, Daily), Daily alcohol consumption, Age quit drinking, Age started drinking, History of drunkenness, Type of alcohol (Liquor, Beer, Wine, Rice wine, Other);

History of exposure to occupational disease hazards.

4. Physical examination

Lips, Dentition, Missing teeth, Decayed teeth, Dentures, Pharynx, Vision in left eye, Vision in right eye, Corrected distance vision in left eye, Corrected distance vision in right eye, Hearing (Can hear, Cannot hear clearly or at all), Motor function (Able to perform smoothly, Unable to perform any movements independently), Fundus examination (Normal, Abnormal), Skin examination (Normal, Flushing, Pallor, Cyanosis, Jaundice, Hyperpigmentation, Other), Sclera examination (Normal, Jaundice, Hyperemia, Other), Lymph node examination, Lung examination, Heart rate, Heart rhythm, Cardiac murmur, Abdominal examination, Edema in lower limbs (None, Unilateral, Bilateral asymmetrical, Bilateral symmetrical), Digital rectal examination (No abnormalities, Tenderness, Mass, Prostate abnormalities, Other), Breast examination, Vulva examination, Vaginal examination, Cervical examination, Uterine examination, Adnexal examination

5. Biochemical tests

Hemoglobin, White blood cells, Platelets, Urine protein, Urine glucose, Urine ketones, Urine occult blood, Fasting blood glucose, Glycated hemoglobin, Hepatitis B surface antigen test, Serum alanine aminotransferase (ALT), Serum aspartate aminotransferase (AST), Albumin, Total bilirubin, Conjugated bilirubin, Serum creatinine, Blood urea, Blood potassium concentration, Blood sodium concentration, Total cholesterol, Triglycerides, Serum low-density lipoprotein cholesterol, Serum high-density lipoprotein cholesterol.

6. Instrumental examinations

Electrocardiogram, Chest X-ray, Abdominal ultrasound, Prostate/bladder ultrasound, Breast ultrasound, Cervical smear.

7.Health evaluation and Guidance

Existing major health issues (Cerebrovascular diseases, Kidney diseases, Heart diseases, Vascular diseases, Eye diseases, Neurological diseases, Other systemic diseases), Health evaluation (No abnormalities, Abnormalities detected, Specific abnormalities), Health guidance (Incorporate into chronic disease management, Recommend re-examination, Recommend referral), Risk factor control recommendations (Quit smoking, Healthy alcohol consumption, Diet, Exercise, Weight loss, Vaccination, Other).

| |
|---|
| **Part 3. Inpatients records** |
| 1.Basic information: |
| Home address, Age, Sex, Place of birth, Current address, Blood type, Drug allergy history. |
| 2.Disease condition: |
| Admission date, Department of admission, Hospital of admission, Discharge date, Discharge status (Transferred, Discharged, Deceased) |
| 3.Diagnosis: |
| Outpatient diagnosis (ICD-10 code), Inpatient diagnosis (ICD-10 code), Pathological diagnosis, External cause of injury or poisoning code (ICD-10 code) |
| **Part 4. Mortality data** |
| 1.Basic information: |
| Home address; Sex; Date of birth; Ethnicity; Educational level (Postgraduate, Bachelor's degree, Associate degree or specialized college, Secondary specialized school, Technical school, High school, Junior high school, Primary school, Illiterate or semi-literate, Unknown);Occupation (Officials of state organs, party and mass organizations, enterprises, and institutions; Professional technicians; Clerks and related personnel; Commercial and service personnel; Agricultural, forestry, animal husbandry, fishery, and water conservancy production personnel; Production and transportation equipment operators and related personnel; Other occupations difficult to classify; Unemployed);Marital status (Single, Married, Widowed, Divorced, Marital status not stated);Medical expense payment method (Urban employee basic medical insurance, Urban resident basic medical insurance, New rural cooperative medical system, Poverty relief, Commercial medical insurance, Full public funding, Full self-pay, Other) |
| 2. Death-related information |
| Personal basic information (same as health record section) |
| **Part 5. Meteorological Data** |
| 1.Temperature: |
| monthly average temperature, maximum temperature, monthly minimum temperature |
| 2.Air conditions: |
| PM2.5, PM2.5 components ($SO_4^{2-}$, $NO_3^-$, $NH_4^+$, BC, OM), $O_3$ |
| 3.Other: |
| nighttime light, normalized difference vegetation index(NDVI) |

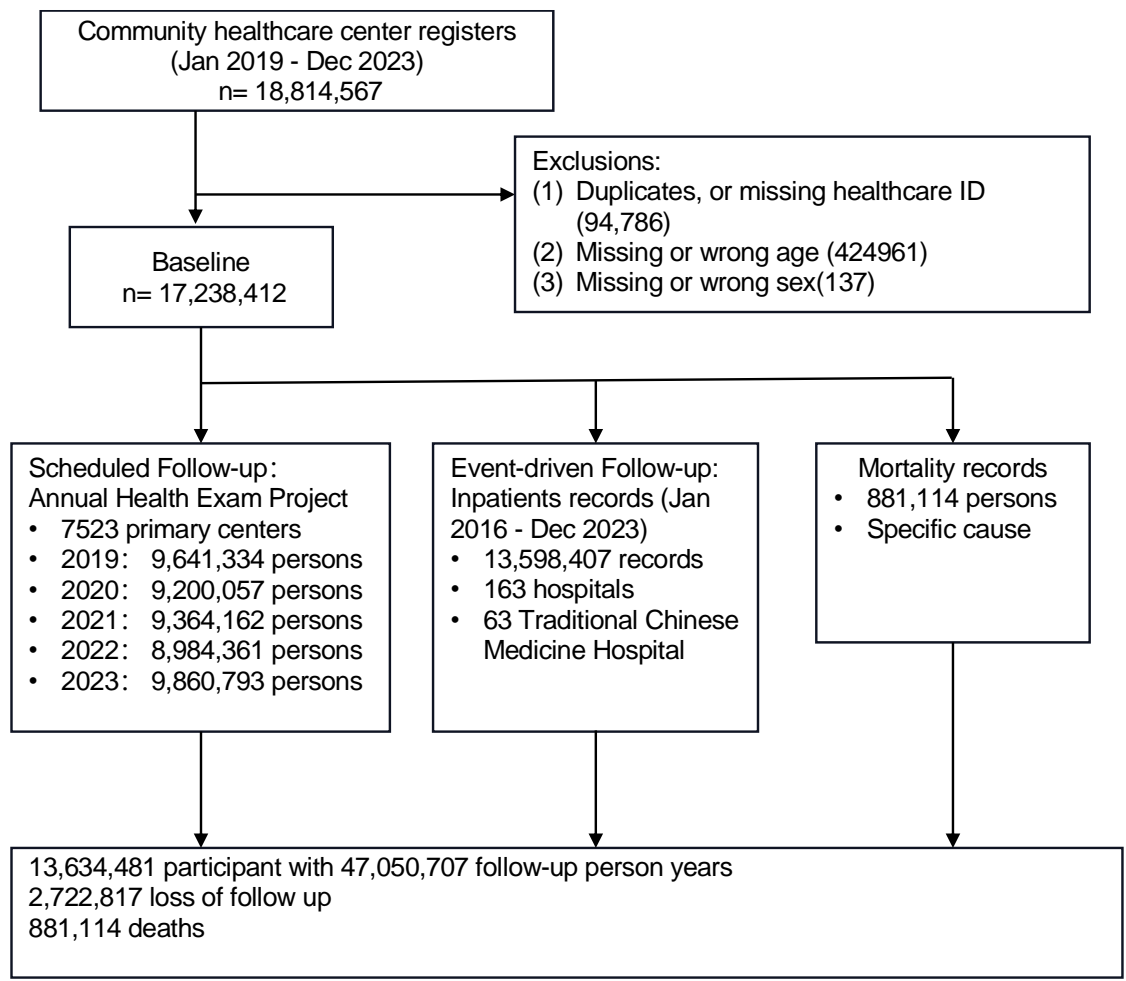

**Figure 1. Cohort construction and follow-up for the Northwest China Real World Cohort.**

Initially comprising 18,814,567 registrants from community healthcare centers, the cohort was refined to 17,238,412 participants after applying exclusion criteria for duplicates, healthcare ID verification, and demographic accuracy (age and sex). Longitudinal health data were systematically collected through annual health examinations at 7,523 primary healthcare facilities, with participation ranging from 9,641,334 to 9,860,793 individuals annually from 2019 through 2023. Event-driven data from inpatient records were accrued from 163 hospitals, including 63 Traditional Chinese Medicine (TCM) hospitals, totaling 13,598,407 records from January 2016 to December 2023. Mortality data with cause-specific details were also integrated, documenting 881,114 deaths. The cohort accumulated 47,050,707 person-years of follow-up, with a loss of follow-up noted for 2,722,817 individuals.

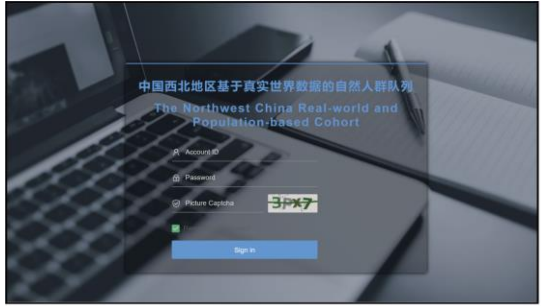
Step 1: Log in

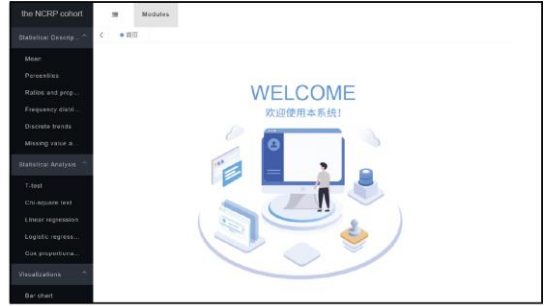
Step 2: Modules selection

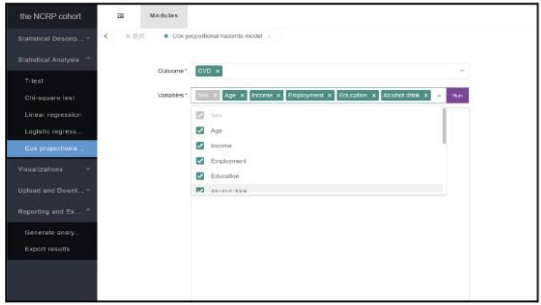
Step 3: Statistic analysis

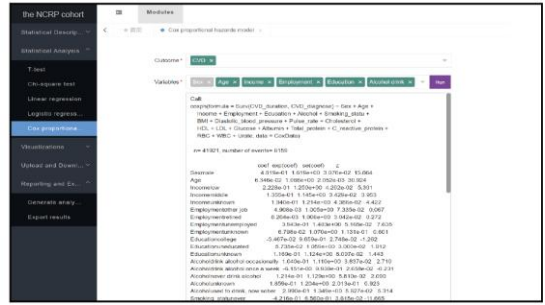
Step 4: Outcome print

**Figure 2. Accessing and analyzing the Northwest China Real World Cohort online.** Upon receiving account credentials and the login address via email, users can directly access the cohort website (Step 1). On the main interface, users can select the relevant functions for data statistics and analysis. Currently, this module consists of five sections: Statistical Description, Statistical Analysis, Visualizations, Upload and Download, and Reporting and Export (Step 2). For instance, if a user chooses to perform a Cox proportional hazards model analysis (Step 3), they can select the disease outcome of interest (e.g., cardiovascular disease) and covariates (e.g., age, sex, occupation, education, and relevant blood test indicators). After clicking RUN, the analysis will be executed, and the results preview will be displayed in the output box below (Step 4).

**Supernumerary table 1. Annual Health Exam Project**

Patient ID ______________ Examiner _____________ Date____________

| Items | PROJECTS |
|---|---|
| **Part 1. General health status** | |
| Current symptoms:__________ 1.None, 2.Headache, 3.Dizziness, 4.Palpitations, 5.Chest tightness, 6.Chest pain, 7.Chronic cough, 8.Sputum production, 9.Shortness of breath, 10.Excessive thirst, 11.Frequent urination, 12.Weight loss, 13.Fatigue, 14.Joint swelling and pain, 15.Blurred vision, 16.Numbness in hands or feet, 17.Urgency of urination, 18.Painful urination, 1.Constipation, 20.Diarrhea, 21.Nausea and vomiting, 22.Dizziness, 23.Tinnitus, 24.Breast tenderness, 25.Other_______ | |
| Temperature °C | Pulse Rate beats/min |
| Respiratory rate beats/min | Blood Pressure — Left ___/___ mmHg ; Right ___/___ mmHg |
| Height cm | Weight kg |
| Waist Circumference cm | Body Mass Index (BMI) Kg/m$^2$ |
| **Part 2. Elderly health self-assessment** | |
| Elderly health self-assessment 1.Satisfied, 2.Basically satisfied, 3.Unclear, 4.Not very satisfied, 5.Dissatisfied | |
| Elderly Self-assessment of ability to perform daily activities * Independent [0–3 points], Mild dependence [4–8 points], Moderate dependence [9–18 points], Unable to perform daily activities [≥19 points] | |
| Elderly Cognitive function **1. Negative preliminary screening** **2. Positive preliminary screening, total score from a simple mental state examination** | |
| Elderly Emotional state Emotional state* **1. Negative preliminary screening** **2. Positive preliminary screening, total score from a depression rating scale for the elderly** | |
| **Part 3. Lifestyle** | |
| Exercise | Frequency:____________ 1.Daily, 2.More than once a week, 3.Occasionally, 4.No exercise |
| | Duration of each exercise session: ___ min \| Years of consistent exercise: ___ years |
| | Type of exercise:______________________ |

| Diet | 1.Balanced diet, 2Mainly meat, 3.Mainly vegetarian, 4. Salt preference, 5.Oil preference, 6.Sugar preference | |
|---|---|---|
| Smoking | Smoking status:______ 1.Never smoked, 2.Quit smoking, 3.Currently smoking | |
| | Daily smoking quantity:______ | |
| | Age started smoking:______ | Age quit smoking:______ |
| Drink | Frequency:______ 1.Never, 2.Occasionally,3. Regularly,4. Daily | |
| | Daily alcohol consumption:______ | |
| | Quit drinking or not:______ 1.Not Quit  2.Quit,Age quit drinking years:______ | |
| | Age started drinking:______(year) | |
| | History of drunkenness in the past year:______  1.Yes 2.No | |
| | Type of alcohol:______ 1.Liquor, 2.Beer, 3.Wine,4. Rice wine,5. Other______ | |
| History of exposure to occupational disease hazards | History of exposure to occupational disease hazards:____ 1 None 2 Yes (work type:    work duration:    years   ) | |
| | Types of Toxic Substances:______ 1.Dust 2.Radioactive Substances 3.Physical Factors 4.Chemical Substances 5.Other | |
| | Protective Measures:______ 1.None 2.Have | |

**Part 4. Physical examination**

| Oral Cavity | Lips:____ 1. Healthy pink, 2. Pale, 3. Cyanosis, 4. Chapped, 5. Herpes |
|---|---|
| | Teeth:____ 1. Normal, 2. Missing teeth, 3. Caries, 4. Dentures (false teeth) |
| | Throat:____ 1. No congestion, 2. Congested, 3. Lymph follicle hypertrophy |
| Vision | Left eye: ____ Right eye: ____ (Corrected vision: Left eye: ____ Right eye: ____) |
| Hearing | 1.Can hear<br>2.Cannot hear clearly or cannot hear at all |
| Motor Function | 1.Can complete movements<br>2.Cannot complete any movements independently |
| Fundoscopy (Eye Exam)* | 1.Normal 2.Abnormal |
| Skin | 1.Normal,2.Flushed,3.Pale,4.Cyanosis,5.Jaundice,6.Hyperpigmentation,7.Others |
| Sclera | 1.Normal 2.Jaundiced 3.Congested 4.Others: |
| Lymph Nodes | 1.Not palpable 2.Supraclavicular 3.Axillary 4.Others: |
| Lungs | Barrel Chest: 1. No 2. Yes |
| | Breath Sounds: 1. Normal 2. Abnormal |
| | Rales: 1. None 2. Dry rales 3. Wet rales 4. Others: |
| Heart | Heart Rate: ____ beats/min<br>Murmurs: 1. None 2. Present<br>Rhythm: 1. Regular 2. Irregular 3. Atrial fibrillation |
| Abdomen | Tenderness: 1.None 2.Present______ (describe)<br>Masses: 1.None 2.Present______ (describe)<br>Hepatomegaly: 1.None 2.Present______ (describe)<br>Splenomegaly: 1.None 2.Present______ (describe)<br>Shifting dullness: 1.None 2.Present______ (describe) |
| Lower Limb Edema | 1.None 2.Unilateral 3.Bilateral asymmetrical 4.Bilateral symmetrical |
| Dorsalis Pedis Artery Pulse | 1.Not palpable<br>2.Palpable, symmetrical<br>3.Palpable, weaker on the left or absent<br>4.Palpable, weaker on the right or absent |
| Digital Rectal Exam: | 1.No abnormalities 2.Tenderness 3.Masses 4.Prostate abnormalities 5.Others |
| Breast Exam | 1.No abnormalities 2.Mastectomy 3.Abnormal lactation 4.Breast lumps 5.Others |
| Gynecological Exam | External Genitalia: 1.No abnormalities 2.Abnormalities<br>Vagina: 1.No abnormalities 2.Abnormalities<br>Cervix: 1.No abnormalities 2.Abnormalities<br>Uterus: 1.No abnormalities 2.Abnormalities<br>Adnexa: 1.No abnormalities 2.Abnormalities |
| Others:* | |
| **Part 5. Biochemical tests** | |
| Blood Routine Test | Hemoglobin     g/L   White blood cells:    ×10⁹/L<br>Platelets    ×10⁹/L   Others |
| Urine Routine Test | Urine protein     Urine glucose     Urine ketones<br>Urine occult blood     Others |
| Fasting Blood Glucose | ____ mmol/L |
| Electrocardiogram (ECG) | 1.Normal   2.Abnormal__________(describe) |
| Fecal Occult Blood Tes | 1.Negative   2.Positive |
| Fecal Occult Blood Tes | 1.Negative   2.Positive |
| Glycosylated Hemoglobin (HbA1c) | % |
| Hepatitis B Surface Antigen | 1.Negative   2.Positive |
| Liver Function | Serum Alanine Aminotransferase (ALT):     U/L<br>Albumin:     g/L<br>Conjugated Bilirubin:     μmol/L<br>Serum Aspartate Aminotransferase (AST):     U/L<br>Total Bilirubin:     μmol/L |
| Kidney Function | Serum Creatinine     μmol/L<br>Blood Potassium     mmol/L<br>Blood Urea Nitrogen (BUN)     mmol/L<br>Blood Sodium     mmol/L |
| Blood Lipids | Total Cholesterol(TC)     mmol/L<br>Triglycerides(TG)     mmol/L<br>Serum Low-Density Lipoprotein Cholesterol (LDL-C)     mmol/L<br>Serum High-Density Lipoprotein Cholesterol (HDL-C)     mmol/L |
| **Part 6. Instrumental examinations** | |
| Chest X-Ray | 1.Normal   2.Abnormal__________(describe) |
| Ultrasound (B-Scan) | Abdominal Ultrasound: 1.Normal   2.Abnormal_____ (describe) |
| | Other     1.Normal   2.Abnormal __________ (describe) |
| Cervical Smear | 1.Normal   2.Abnormal __________ (describe) |
| Others | |
| **Part 7. Health Evaluation and Guidance** | |
| Cerebrovascular Diseases | 1.None found 2.Ischemic stroke 3.Cerebral hemorrhage 4.Subarachnoid hemorrhage 5.Transient ischemic attack 6.Other: |
| Kidney Diseases | 1.None found 2.Diabetic nephropathy 3.Renal failure 4.Acute nephritis 5.Chronic nephritis 6.Other: |
| Heart Diseases | 1.None found 2.Myocardial infarction 3.Angina pectoris 4.Coronary artery bypass grafting (CABG) 5.Congestive heart failure 6.Precordial pain 7.Other |
| Vascular Diseases | 1.None found 2.Aortic dissection 3.Arterial occlusive disease 4.Other |
| Eye Diseases | 1.None found 2.Retinal hemorrhage or exudation 3.Papilledema 4.Cataract 5.Other: |
| Neurological Diseases | 1.None found 2.Present___________(describe) |
| Other Systemic Diseases | 1.None found 2.Present___________(describe) |

1.No abnormalities detected in the physical examination

2.Abnormalities detected:

Abnormality 1: _____

Abnormality 2: _____

Abnormality 3: _____

Abnormality 4: _____

| | |
|---|---|
| 1.Included in chronic disease management  2.Recommended follow-up  3.Recommended referral | Risk Factor Control  1.Quit smoking  2.Healthy drinking  3.Diet  4.Exercise  5.Weight loss (Target: _____ Kg)  6.Recommended vaccination  7.Other: _____ |
| 1.Included in chronic disease management  2.Recommended follow-up  3.Recommended referral | Risk Factor Control  1.Quit smoking  2.Healthy drinking  3.Diet  4.Exercise  5.Weight loss (Target: _____ Kg)  6.Recommended vaccination  7.Other: _____ |

**Form Instructions:**

1. This form is used for annual health check-ups of the elderly, patients with hypertension, type 2 diabetes, and severe mental disorders. It may also be used for general population health check-ups. However, it is not required for tuberculosis patients, pregnant women, or children aged 0-6 years.

**2. General Condition:**

- Body Mass Index (BMI) = weight (kg) / height squared (m²).

- **Self-Care Ability Assessment for the Elderly:** This item must be completed by people aged 65 and above. Please refer to the appendix of the elderly health management service specification (Supernumerary table 2).

- **Cognitive Function Screening for the Elderly:** Tell the examinee, "I am going to say the names of three objects (e.g., pencil, truck, book), please repeat them immediately." After one minute, ask the examinee to repeat the three object names again. If the examinee cannot immediately repeat them or cannot recall the three names after one minute, the result is considered positive, requiring further examination with the "Mini-Mental State Examination (MMSE)" scale (Supernumerary table 3).

- **Emotional State Screening for the Elderly:** Ask the examinee, "Do you often feel sad or depressed?" or "How is your mood?" If the answer is "Yes" or "I think it's not very good," the result is considered positive, requiring further examination using the "Geriatric Depression Scale (GDS) (Supernumerary table 4)."

3. **Lifestyle:**

- **Physical Exercise:** This refers to intentional exercise activities aimed at fitness. It does not include activities required for work or other needs, such as cycling to work or performing heavy physical labor. The most commonly used exercise method should be specified.

- **Smoking Status:** Those who have never smoked do not need to fill in the "daily cigarette consumption," "starting age of smoking," or "quitting age." Those who have quit smoking should fill in the relevant information about their smoking habits before quitting.

- **Alcohol Consumption:** Those who never drink do not need to fill in any other alcohol-related fields. Those who have quit drinking should fill in the relevant information before quitting. The daily alcohol consumption should be converted to equivalent white spirit amounts (Beer/10 = white spirit amount, Red wine/4 = white spirit amount, Yellow wine/5 = white spirit amount).

- **Occupational Exposure:** This refers to contact with chemicals, toxins, or radiation due to the patient's occupation. If present, specify the chemical, toxin, or radiation source, or fill in "unknown."

- **History of Exposure to Occupational Disease Risk Factors:** This refers to exposure to dust, radioactive materials, physical factors, or chemical substances due to the patient's occupation. If present, specify the specific dust, radioactive material, physical factors, or chemical substances, or fill in "unknown."

**4. Organ Function:**

- **Vision:** Record the specific value measured using the logarithmic vision chart (five-point scale). For those wearing glasses, measure the corrected vision while wearing their usual glasses.

- **Hearing:** Whisper "What is your name?" next to the examinee's ear (ensure the examiner's face is out of the examinee's sight), and assess the hearing ability.

- **Motor Function:** Ask the examinee to perform the following actions: "Touch the back of your head with both hands," "Pick up this pen," "Stand up from the chair, walk a few steps, turn around, and sit down." These actions help assess the motor function.

**5. Physical Examination:**

If any abnormalities are found, specify them on the lines provided, such as the location and number of palpable lymph nodes, heart murmurs, or the size of the liver or spleen upon palpation. It is recommended that regions capable of conducting fundoscopy, especially for hypertensive or diabetic patients, perform the test.

- **Fundoscopy:** If abnormalities are found, describe the specific results.
- **Dorsalis Pedis Artery Pulse:** Diabetic patients must undergo this test.
- **Breast:** Examine for any abnormalities in appearance, abnormal lactation, or lumps.
- **Gynecology:**
    - **External Genitalia:** Record the development and reproductive status (e.g., unmarried, married without children, or multipara). If abnormalities are found, provide a detailed description.
    - **Vagina:** Record whether it is patent, the condition of the mucous membrane, the quantity, color, and characteristics of the discharge, and whether there is an odor.
    - **Cervix:** Record the size, texture, and whether there is erosion, tearing, polyps, or glandular cysts; and whether there is contact bleeding or tenderness.
    - **Uterus:** Record the position, size, texture, mobility, and whether there is tenderness.
    - **Adnexa:** Record whether there are any masses, thickening, or tenderness. If a mass is palpable, describe its position, size, texture, surface smoothness, mobility, tenderness, and relationship with the uterus and pelvic wall. Record findings for both sides separately.

**6. Auxiliary Examinations:**

This item is conducted based on local circumstances and for specific populations. Free auxiliary examinations for elderly patients, hypertensive patients, type 2 diabetes patients, and those with severe mental disorders are implemented according to the corresponding service specifications.

- **Urine Routine:** Record "negative" as "–," and "positive" based on the results as "+," "++," "+++," or "++++," or fill in quantitative results with measurement units.
- **Fecal Occult Blood, Liver Function, Kidney Function, Chest X-ray, and Ultrasound (B-Scan) Results:** If abnormalities are found, provide a detailed description of the abnormal results. For B-scan ultrasound, specify the examined area. Abdominal B-scan ultrasound is a free examination item for elderly individuals aged 65 and above.
- **Other:** Record any other auxiliary examination results not listed in the table under "Other."

**7. Existing Major Health Problems:**
This refers to diseases that have appeared or are continuously present and affect current health conditions. Multiple selections are allowed. If hypertension, diabetes, or other existing diseases are present or new diseases have been diagnosed, they should also be recorded in the medical history section of the individual's basic information.

**8 Health Evaluation:**
No abnormalities mean no newly diagnosed diseases, and existing diseases are well controlled without worsening or progression. If abnormalities are present, specify them, including conditions such as hypertension, diabetes, impaired self-care, and psychological screening abnormalities.

**9. Health Guidance:**
Chronic disease management includes regular follow-ups and health check-ups for key populations such as those with hypertension, diabetes, and severe mental disorders. The weight loss target refers to the specific weight reduction goal that should be achieved before the next health check-up, based on the individual's condition.

**Supernumerary table 2. Elderly Self-Care Ability Assessment Form**

This is a self-assessment form. The evaluation is based on five aspects as outlined in the table below. After summing the scores from each aspect, the results are interpreted as follows: 0 to 3 points: Independent, 4 to 8 points: Mild dependence, 9 to 18 points: Moderate dependence, ≥19 points: Unable to care for oneself.

| Assessment Items Content and Scoring | Level of Dependence | | | | Scoring |
|---|---|---|---|---|---|
| | **Independent** | **Mild dependence** | **Moderate dependence** | **Unable to care for oneself** | |
| **Eating:** Activities such as using utensils to bring food to the mouth, chewing, and swallowing. | **Independently completed** | - | **Requires assistance, such as cutting or mixing food** | **Completely needs assistance;** | |
| **Scoring** | **0** | **0** | **3** | **5** | |
| **Grooming:** Activities such as combing hair, washing the face, brushing teeth, shaving, and bathing. | **Independently completed** | **Can independently wash hair, comb hair, wash face, brush teeth, and shave; requires assistance for bathing** | **With assistance and within an appropriate amount of time, can complete part of the grooming activities.** | **Completely needs assistance;** | |
| **Scoring** | **0** | **1** | **3** | **7** | |
| **Dressing:** Activities such as putting on clothes, pants, socks, and shoes | **Independently completed** | - | **Requires assistance and can complete part of the dressing within an appropriate time;** | **Completely needs assistance;** | |
| **Scoring** | **0** | **0** | **3** | **5** | |
| **Toileting:** Activities related to urination, defecation, and self-control. | **No assistance needed, self-controllable** | **Occasional incontinence, but generally able to use the toilet or a bedpan** | **Frequent incontinence, but can still use the toilet or bedpan with frequent reminders and assistance;** | **complete incontinence, fully requires assistance** | |
| **Scoring** | **0** | **0** | **5** | **10** | |
| **Mobility:** Activities such as standing, walking indoors, climbing stairs, and outdoor activities. | **Independently completes all activities** | **Can complete standing, walking, and climbing stairs with minor assistance or the use of assistive devices** | **Needs significant physical assistance to stand and walk, unable to climb stairs.** | **Bedridden, all activities require full assistance** | |
| **Scoring** | **0** | **0** | **5** | **10** | |
| **Total score** | | | | | |

# Supernumerary table 3. The Mini-Mental State Exam

Patient ID ______________ Examiner _____________ Date____________

| Items | Describe | Maximum score | Scoring |
|---|---|---|---|
| Orientation | What is the (year) (season) (date) (day) (month)? | 5 | |
| | Where are we (state) (country) (town) (hospital) (floor)? | 5 | |
| Registration | Name 3 objects: 1 second to say each. Then ask the patient. all 3 after you have said them. Give 1 point for each correct answer. Then repeat them until he/she learns all 3. Count trials and record. | 3 | |
| Attention and Calculation | Serial 7's. 1 point for each correct answer. Stop after 5 answers. | 5 | |
| Recall | Ask for the 3 objects repeated above. Give 1 point for each correct answer | 3 | |
| Language | Name a pencil and watch | 2 | |
| | Repeat a sentence | 1 | |
| | Follow a 3-stage command:"Take a paper in your hand, fold it in half, and put it on the floor." | 3 | |
| | Read and obey the following: CLOSE YOUR EYES | 1 | |
| | Write a sentence. | 1 | |
| | Copy the design shown. 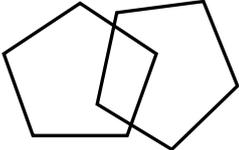 | 1 | |
| | Total score | 30 | |

# Supernumerary table 4. Geriatric Depression Scale

Patient ID _______________ Examiner _______________ Date_____________

**Instructions:** Choose the answer that best reflects how you have felt over the past week. (Yes, No)

1. Are you generally satisfied with your life? (No)
2. Have you dropped many of your activities and interests? (Yes)
3. Do you feel your life is empty? (Yes)
4. Do you often feel bored? (Yes)
5. Do you feel hopeful about the future? (No)
6. Are you bothered by thoughts you can't get out of your head? (Yes)
7. Do you feel full of energy most of the time? (No)
8. Are you afraid something bad is going to happen to you? (Yes)
9. Do you feel happy most of the time? (No)
10. Do you often feel helpless? (Yes)
11. Do you frequently feel restless and can't sit still? (Yes)
12. Do you prefer to stay at home rather than doing new things? (Yes)
13. Do you frequently worry about the future? (Yes)
14. Do you feel your memory is worse than it used to be? (Yes)
15. Do you find life to be wonderful now? (No)
16. Do you often feel downhearted and blue? (Yes)
17. Do you feel that your life is meaningless as it is now? (Yes)
18. Do you often feel troubled by things from the past? (Yes)
19. Do you find life exciting? (No)
20. Is it hard for you to start new projects? (Yes)
21. Do you feel full of life? (No)
22. Do you feel that your situation is hopeless? (Yes)
23. Do you feel that most people are better off than you? (Yes)
24. Do you often get upset over little things? (Yes)
25. Do you frequently feel like crying? (Yes)
26. Do you have trouble concentrating? (Yes)
27. Do you feel cheerful in the morning? (No)
28. Do you avoid social gatherings? (Yes)
29. Is it easy for you to make decisions? (No)
30. Is your mind as clear as it used to be? (No)

**Scoring:**

Each response in parentheses after the question indicates depression, and a matching answer receives 1 point.

- **0-10 points:** Normal
- **11-20 points:** Mild depression
- **21-30 points:** Moderate to severe depression

**Total score:**_____________